\documentclass[iop]{emulateapj}
\usepackage[pdftex]{color}
\pdfoutput=1

\newcommand\6{{\footnotesize VI}}
\newcommand\5{{\footnotesize V}}
\newcommand\4{{\footnotesize IV}}
\newcommand\3{{\footnotesize III}}
\newcommand\2{{\footnotesize II}}
\newcommand\1{{\footnotesize I}}
\newcommand\lam{{$\lambda$}}

\newcommand\kms{$\rm{km s^{-1}}$}

\newcommand\pd{\phantom{$-$}}

\lefthead{Evans et al.}
\righthead{A massive runaway star from 30 Dor}

\shorttitle{A massive runaway star from 30 Dor}
\shortauthors{Evans et al.}

\begin{document}
\bibliographystyle{plainnat}

%% LaTeX will automatically break titles if they run longer than
%% one line. However, you may use \\ to force a line break if
%% you desire.

\title{A massive runaway star from 30~Doradus\altaffilmark{*}}

\slugcomment{Received 2010 February 24; Accepted 2010 April 8}

%% Use \author, \affil, and the \and command to format
%% author and affiliation information.
%% Note that \email has replaced the old \authoremail command
%% from AASTeX v4.0. You can use \email to mark an email address
%% anywhere in the paper, not just in the front matter.
%% As in the title, you can use \\ to force line breaks.

\author{C. J. Evans\altaffilmark{1}, 
N. R. Walborn\altaffilmark{2}, 
P. A. Crowther\altaffilmark{3}, 
V. H\'{e}nault-Brunet\altaffilmark{4}, 
D. Massa\altaffilmark{2},\\ 
W. D. Taylor\altaffilmark{4}, 
I. D. Howarth\altaffilmark{5}, 
H. Sana\altaffilmark{6,7}, 
D. J. Lennon\altaffilmark{2,8}, 
and J. Th. van Loon\altaffilmark{9}
}

\altaffiltext{*}{Based on observations obtained at the European
  Southern Observatory Very Large Telescope in program 184.D-0222 and
  with the NASA/ESA {\em Hubble Space Telescope} in program 11484,
  obtained at the Space Telescope Science Institute, which is operated
  by the Association of Universities for Research in Astronomy, Inc.,
  under NASA contract NAS 5-26555.}
\altaffiltext{1}{UK Astronomy Technology Centre, 
                 Royal Observatory Edinburgh, 
                 Blackford Hill, 
                 Edinburgh, 
                 EH9 3HJ, UK}
\altaffiltext{2}{Space Telescope Science Institute, 
                 3700 San Martin Drive,
                 Baltimore, 
                 MD 21218, USA} 
\altaffiltext{3}{Department of Physics and Astronomy, 
                 University of Sheffield, 
                 Hounsfield Road, 
                 Sheffield, 
                 S3~7RH, UK}
\altaffiltext{4}{Scottish Universities Physics Alliance (SUPA),
                 Institute for Astronomy, 
                 University of Edinburgh, 
                 Royal Observatory Edinburgh, 
                 Blackford Hill, 
                 Edinburgh, 
                 EH9 3HJ, UK}
\altaffiltext{5}{Department of Physics and Astronomy, 
                 University College London, 
                 Gower Street, 
                 London, 
                 WC1E~6BT, UK}
\altaffiltext{6}{European Southern Observatory, 
                 Alonso de Cordova 1307, 
                 Casilla 19001, 
                 Santiago 19, Chile}
\altaffiltext{7}{Sterrenkundig Instituut Anton Pannekoek, 
                 Universiteit van Amsterdam, 
                 Postbus 94249, 
                 1090 GE Amsterdam, 
                 The Netherlands}
\altaffiltext{8}{European Space Agency,
                 Research and Scientific Support Department, 
                 3700 San Martin Drive,
                 Baltimore, 
                 MD 21218, USA} 
\altaffiltext{9}{Astrophysics Group, 
                  School of Physical and Geographical Sciences, 
                  Keele University, 
                  Staffordshire, ST5 5BG, UK}

\begin{abstract}
  We present the first ultraviolet (UV) and multi-epoch optical
  spectroscopy of 30~Dor~016, a massive O2-type star on the
  periphery of 30~Doradus in the Large Magellanic Cloud.  The UV data
  were obtained with the Cosmic Origins Spectrograph on the {\em
    Hubble Space Telescope} as part of the Servicing Mission
  Observatory Verification program after Servicing Mission~4, and
  reveal \#016 to have one of the fastest stellar winds known.  From
  analysis of the C~\4 \lam\lam1548-51 doublet we find a terminal
  velocity, $v_\infty$\,$=$\,3450\,$\pm$\,50\,\kms.  Optical
  spectroscopy is from the VLT-FLAMES Tarantula Survey, from which we
  rule out a massive companion (with 2\,d\,$<$\,$P$\,$<$\,1\,yr) to
  a confidence of 98\%.  The radial velocity of \#016 is offset from the
  systemic value by $-$85\,\kms, suggesting that the star has 
  traveled the 120\,pc from the core of 30 Doradus as a runaway, 
  ejected via dynamical interactions.
\end{abstract}


\section{Introduction}\label{intro}

30~Doradus in the Large Magellanic Cloud (LMC) is the richest 
H\,\2 region in the Local Group, providing an excellent template
with which to study regions of intense star formation, and both
stellar and cluster evolution.  It harbors a significant fraction of
the most massive and luminous stars known, with a rich population of
the earliest O-type stars (e.g., Melnick 1985; Walborn \& Blades
1997), particularly in R136, the dense cluster at its core (Massey \&
Hunter 1998).

The O3 spectral class was introduced by Walborn (1971) to accommodate
stars in which He~\1 \lam4471 was absent in moderate-resolution
photographic spectra, compared to the very weak absorption seen in
O4-type spectra.  The classification scheme was extended further by
Walborn et al. (2002) to include the new types of O2 and O3.5 to
delineate the behavior of the N~\3, N~\4, and N~\5 features seen in
the earliest types. Even in the age of large multi-object surveys only
a few tens of stars are known with O2--O3.5 types.  Although very rare,
their influence is far-reaching as they are expected to evolve
rapidly into nitrogen rich Wolf--Rayet stars (WN types), plausible
progenitors of supernovae and, potentially, gamma-ray bursts (Smartt
2009).

\begin{figure}
\begin{center}
\includegraphics[width=8cm]{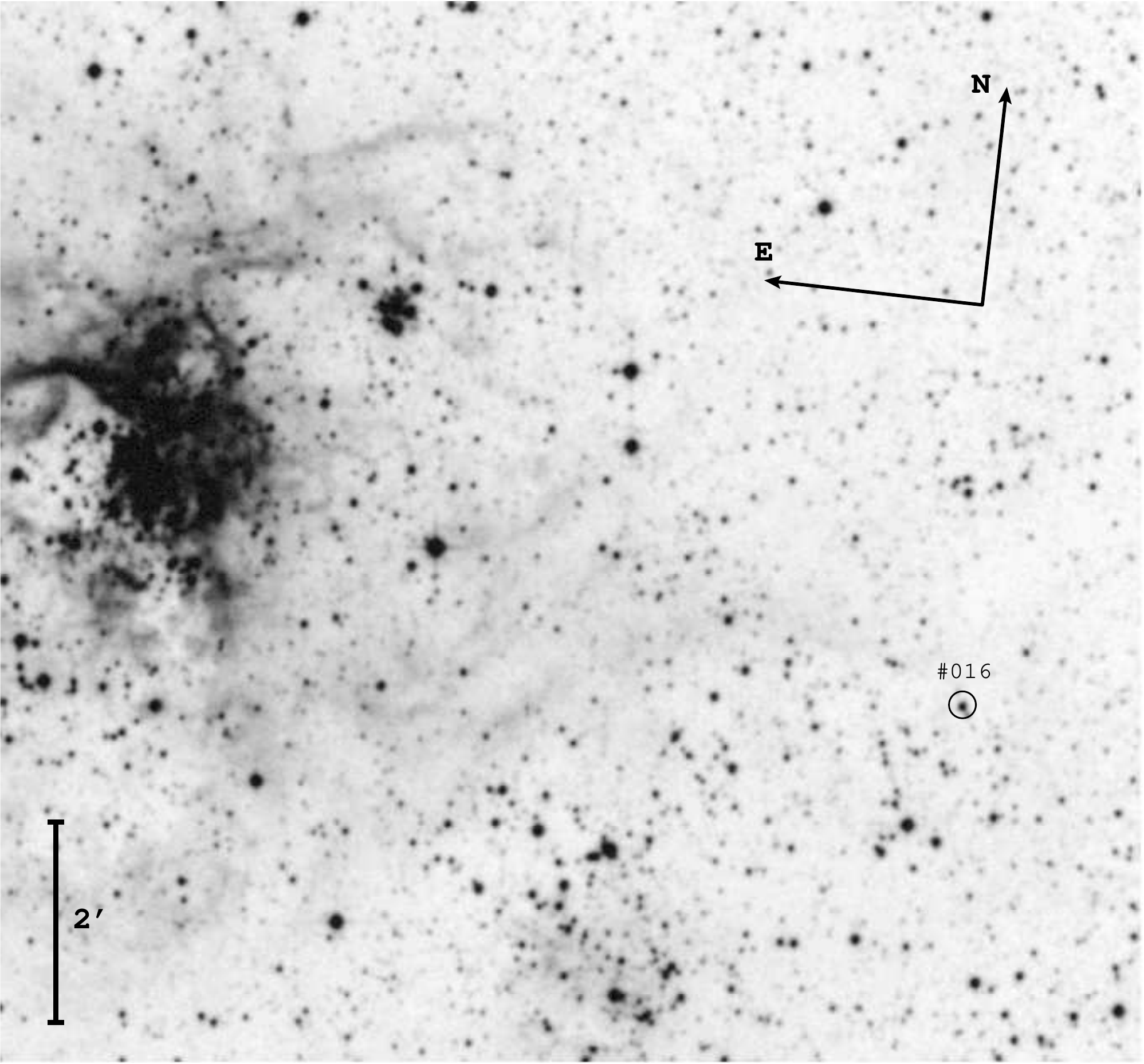}
\caption{Digital Sky Survey ``IR'' ($\sim${\em I}-band) image showing 
the location of 30~Dor~016 (encircled) relative to the 30~Dor complex.}
\label{fig1} 
\end{center}
\end{figure}

Observations with the 2-degree Field (2dF) instrument at the
Anglo-Australian Telescope revealed a new O2-type star on the western
fringes of 30~Doradus (Figure~\ref{fig1}), with a radial velocity of
$\sim$85\,\kms\/ lower than the systemic velocity of nearby massive
stars (e.g., Bosch, Terlevich \& Terlevich 2009).  New multi-epoch
spectroscopy of this star, 30~Dor~016 in the VLT-FLAMES Tarantula
Survey (Evans et al. 2010), now enables us to rule out the presence of
a close massive companion to a high level of confidence, suggesting
that the star might have been ejected from the denser central region.
Here, we combine these observations with new ultraviolet (UV)
spectroscopy of \#016, some of the first data taken with the Cosmic
Origins Spectrograph (COS) on the {\em Hubble Space Telescope (HST)},
which reveal the star to have one of the highest wind terminal
velocities seen to date in any massive star.

\section{Observations}\label{obs}

\subsection{UV spectroscopy}\label{obs_uv}

30~Dor~016 ($\alpha$\,=\,05$^{\rm h}$\,37$^{\rm m}$\,08\fs88, 
$\delta$\,=\,$-$69$^\circ$\,07$'$\,20\farcs36, J2000) was observed 
with {\em HST}-COS as part of the Servicing
Mission Observatory Verification (SMOV) program in 2009 July using the
G130M and G160M gratings in the far-UV channel.  The data were
obtained as part of a focus-check run early in the SMOV phase of COS
operations so we do not have full wavelength coverage as the gratings
were not moved.  Out-of-focus exposures ($<$25\% of the total) were
excluded when co-adding the spectra, with final exposure times of
5671\,s (G130M) and 4511\,s (G160M).  Estimates of the signal-to-noise
ratio (S/N) in the combined spectrum are complicated by the large
number of metallic absorption lines in the spectra of hot stars,
particularly while the instrument is still undergoing full
characterization.  Here, we use selected continuum regions that are
relatively free of lines to estimate the noise, yielding S/N
$\gtrsim$40.

Preliminary characterization of the COS on-orbit performance has found
broad non-Gaussian wings in the line-spread function of the
spectrograph owing to wavefront errors from the primary and secondary
mirrors of the telescope (Ghavamian et al. 2009).
This reduces the detection threshold for weak absorption features and
modifies the profiles of saturated lines.  To account for these
effects, our models of the C~\4 line in Section~\ref{sei} were
convolved by the line-spread functions determined by the COS team.

The COS spectrum of \#016 is shown in Figure~\ref{fig2}, together with
an {\em HST} Space Telescope Imaging Spectrograph (STIS) E140M spectrum
from program 9434 (P.I. Lauroesch) of HDE\,269810
(Sk$-$67$^\circ$~211), classified as O2~III(f$^\ast$) by Walborn et
al. (2002).  They are close twins, but with
subtle differences --- the most striking being the morphology of the
C~\4 absorption and the P~Cygni emission at He~\2 \lam1640.  Indeed,
the C~\4 doublet in \#016 is saturated to a large velocity, indicative
of a very fast stellar wind.

\subsection{Optical data}
Optical spectroscopy of \#016 was obtained as part of two
programs with 2dF, before and after the AAOmega upgrade (Sharp et al.,
2006).  Three 2dF-AAOmega spectra were obtained on 2006 February 22--23
(P.I. JVL), with a 2dF spectrum obtained on 2004 December 8 (P.I. IDH), as
summarized in Table~\ref{obsinfo}.

The star was also observed as part of the Tarantula Survey (full
details to be published elsewhere).  Observations of \#016 with three
of the standard FLAMES settings (LR02, LR03, and HR15N) were obtained
over the period 2008 December 17--22.  Pairs of exposures were taken
in each observing block (OB), with three OBs observed at the LR02 and
LR03 settings, and two at HR15N.  Individual exposure times were 1815s
(LR02, LR03) and 2265s (HR15N).  The spectral coverage and resolution,
as defined by the full width at half-maximum (FWHM) of the arc
calibration lines, is given in Table~\ref{obsinfo}.  

The survey features repeat LR02 observations to detect radial
velocity variables; for \#016 these were obtained on 2009 January 29,
February 28, and October 8 (epochs 4, 5, and 6 in
Table~\ref{obsinfo}, respectively).  The modified Julian Dates (MJD)
for each epoch are listed in Table~\ref{obsinfo}. Each co-added
pair has an S/N\,$>$\,100.

\begin{center}
\begin{deluxetable}{lcccc}
\tabletypesize{\footnotesize}
\tabletypesize{\footnotesize}
\tablewidth{0pc}
\tablecolumns{5}
\tablecaption{Observational epochs and differential radial velocities\label{obsinfo}}
\tablehead{\colhead{Instrument}  &
\colhead{FWHM} & \colhead{Epoch} & \colhead{MJD} 
& \colhead{$\Delta v_{\rm r}$} \\ 
\colhead{and Setting} & \colhead{[\AA]} & \colhead{} & \colhead{} & 
\colhead{[\kms]}}
\startdata
FLAMES-LR02 & 0.61 & 1 & 54817.223 & \pd0.0 \\
\pd{[\lam\lam3960-4564]} & & 2 & 54817.267 & \pd1.7 \\
     & & 3 & 54822.058 & $-$1.8 \\ 
     & & 4 & 54860.105 & $-$1.7 \\
     & & 5 & 54890.041 & $-$1.7 \\
     & & 6 & 55112.361 & $-$2.0  \\
\hline
FLAMES-LR03 & 0.56 & 1 & 54818.243 & \pd0.5 \\
\pd{[\lam\lam4499-5071]} & & 2 & 54818.287 & $-$0.4 \\
     & & 3 & 54818.330 & $-$2.8 \\
\hline
FLAMES-HR15N & 0.41 & 1 & 54818.076 & \pd$-$ \\
\pd{[\lam\lam6442-6817]} & & 2 & 54818.129 & \pd$-$ \\
\hline
2dF-AAOmega & 1.0 & $-$ & 53788.442 & \pd8.8 \\
\pd{[1700B]}& 1.0 & $-$ & 53788.490 & $-$3.0 \\
\pd{[1500V]}& 1.25 & $-$ & 53789.430 & $-$3.6 \\
\hline
2dF [1200B] & 2.75 & $-$ & 53347.564 & \pd$-$
\enddata
\tablecomments{$\Delta v_{\rm r}$ results from cross-correlation with the
first epoch LR02 observations.}
\end{deluxetable}
\end{center}

\subsubsection{Spectral classification}\label{class}

Informed by the apparent absence of detectable radial-velocity variations
(Section~\ref{rv}), the individual LR02 and LR03 spectra were
co-added, then merged in the overlap region, yielding an 
S/N~$>$~350.

\begin{figure*}
\begin{center}
\includegraphics[width=15cm]{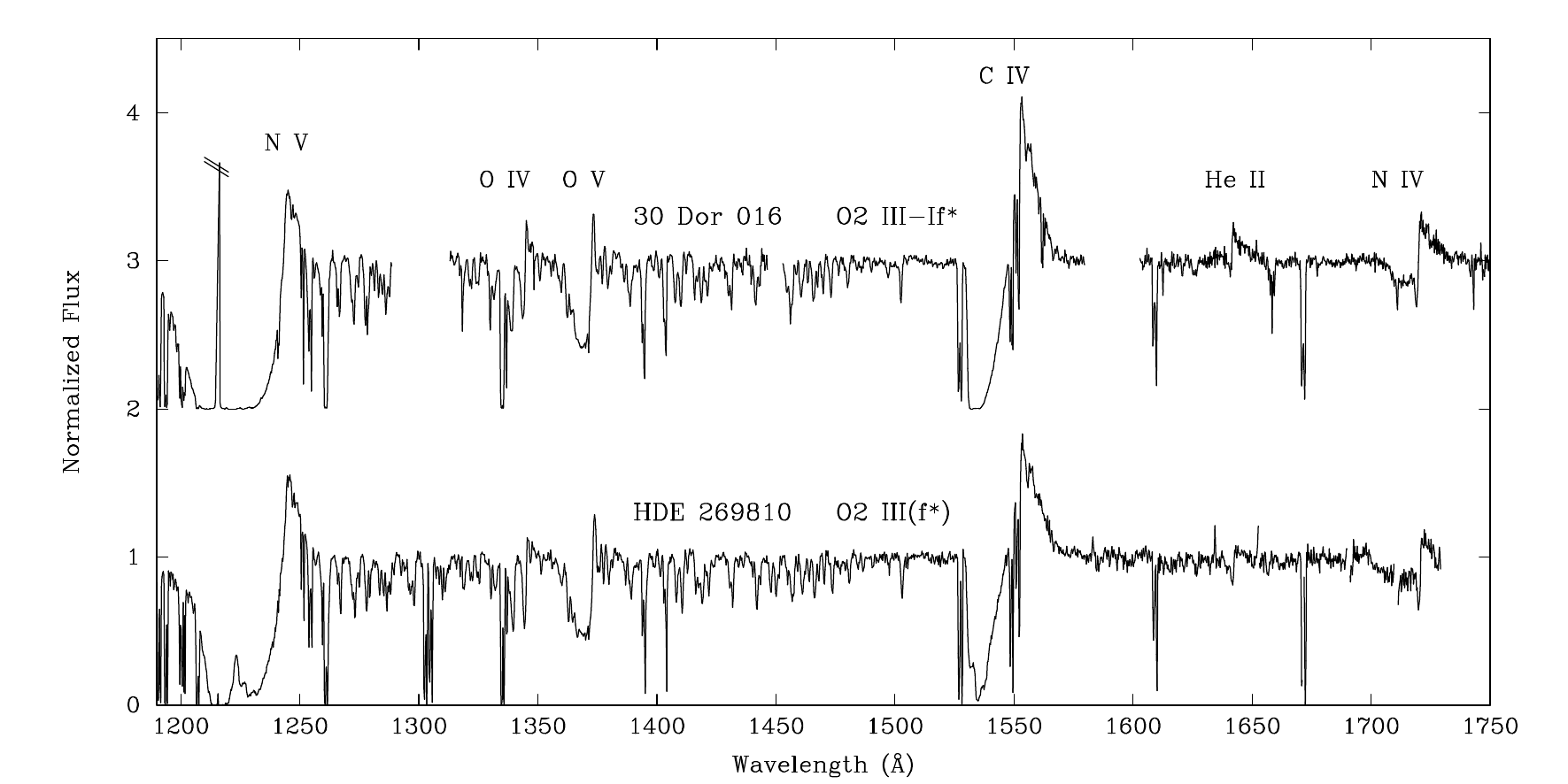}
\includegraphics[viewport=40 90 530 765,scale=0.55,angle=270]{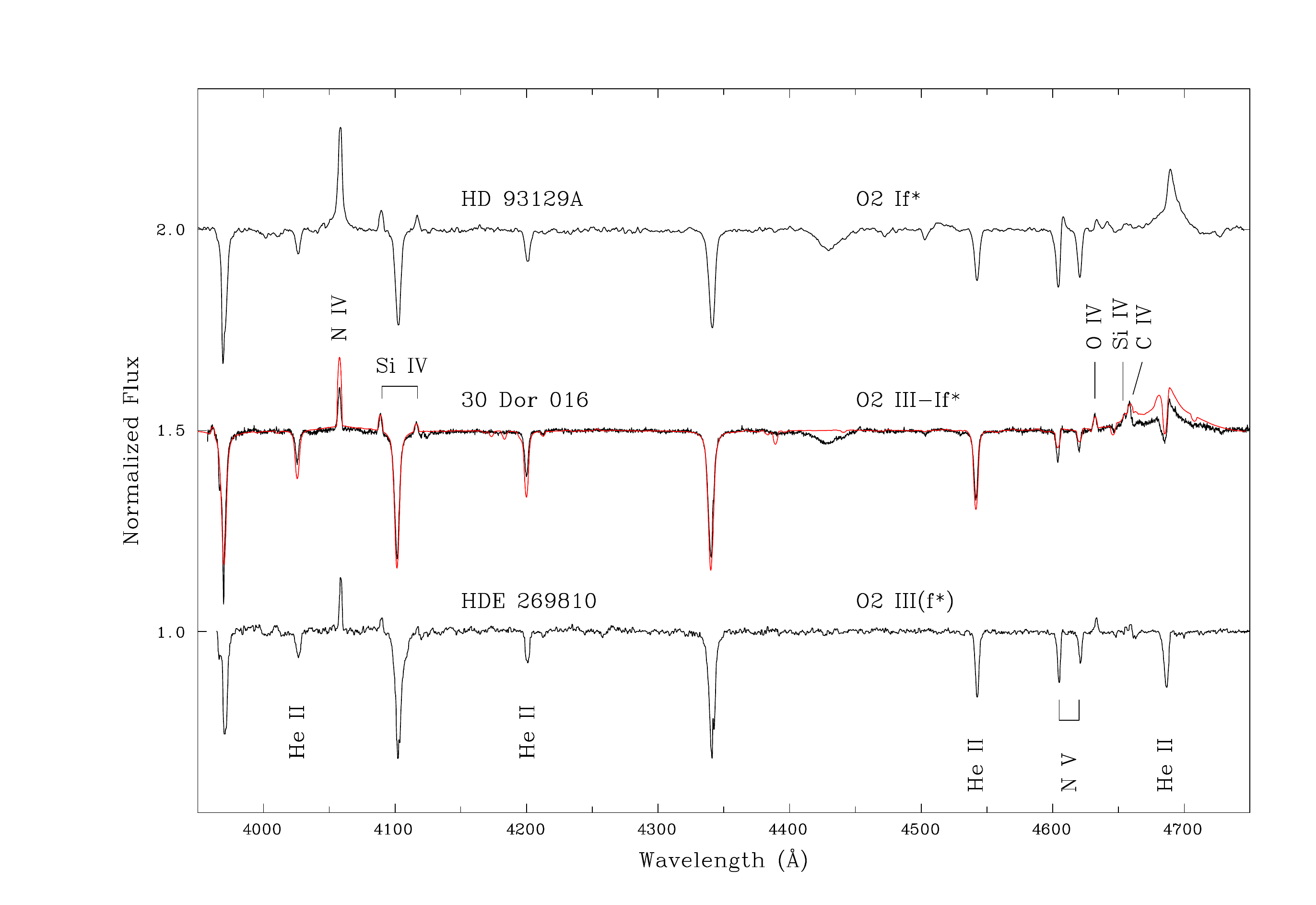}
\caption{{\em Upper panel:} {\em HST}-COS spectrum of 30~Dor 016 and
  an {\em HST}-STIS spectrum of its near twin, HDE\,269810 (both binned
  to 0.25\,\AA).  The wind features identified are N~\5
  \lam\lam1239-43, O~\4 \lam\lam1339-44, O~\5 \lam1371, C~\4
  \lam\lam1548-51, He~\2 \lam1640, and N~\4 \lam1718.  {\em Lower
    panel:} FLAMES spectrum of \#016, illustrating its place in the
  luminosity sequence at type O2 compared to HD\,93129A and
  HDE\,269810.  Emission lines identified in \#016 are N~\4 \lam4058;
  Si~\4 \lam\lam4089, 4116, 4654; O~\4 \lam4632; C~\4 \lam4658.
  Absorption lines identified in HDE\,269810 are He~\2 \lam\lam4026,
  4200, 4542, 4686; N~\5 \lam\lam4604, 4620. The {\sc cmfgen} model
  for \#016 is overplotted in red.}\label{fig2}
\end{center}
\end{figure*}

The \lam\lam3950-4750\,\AA\/ region of \#016 is shown in
Figure~\ref{fig2}, compared with spectra of HDE\,269810 and
HD\,93129A from Walborn et al. (2002).  The most notable feature in
the spectrum of \#016 is the strong N~\4 \lam4058 emission combined
with an absence of N~\3 \lam4640 emission, suggesting that the star is of the
very earliest type in the existing classification framework.  Also
note the O~\4 \lam 4632, Si~\4 \lam4654, and C~\4 \lam4658 emission, and
the He~\2 \lam4686 P~Cygni profile.  Accurate rectification of the
\lam4686 and H$\alpha$ regions is a notorious problem in
observations of O-type stars, but we see broadened wings to the He~\2
line in both manual and scripted rectifications.

As in the UV, the spectrum of \#016 is similar to that of HDE\,269810.
While the He~\2 \lam4686 profile in \#016 suggests a somewhat more
luminous star compared to HDE\,269810, it is not as remarkable as the
emission seen in HD\,93129A (O2~If$^\ast$; Walborn et al. 2002),
leading to our intermediate classification of O2~III-If$^\ast$.

\subsubsection{Stellar radial velocities}\label{rv}

There are relatively few optical lines for radial velocity analysis in
such an early-type spectrum.  The Balmer hydrogen lines and He~\2
\lam4686 are strongly affected by the wind, so do not provide a true
diagnostic of the stellar radial velocity.  Overplotting of the data
from each epoch (including those from 2dF) revealed no significant
velocity shifts nor broadening/asymmetries in the lines, suggesting
that a high-mass companion is not present and that \#016 is a prime
candidate as an ejected, massive runaway from R136.

The {\sc xcorr} routine in the {\sc dipso} spectral analysis package
(Howarth et al. 2004) was used to cross-correlate the co-added pairs
of LR02 spectra with the first epoch of LR02 observations.  The
velocity offsets, $\Delta v_{\rm r}$ (see Table~\ref{obsinfo}), are
found from fits to the cross-correlation functions (CCFs) with
Gaussian profiles generated using the {\sc dipso}
emission-line fitting (ELF) routine.  Although the CCFs are not
Gaussian, the ELF fits were restricted to the range
$-$100\,$<$\,$\Delta v_{\rm r}$\,$<$\,100\,\kms, yielding robust
estimates with fitting errors of 3--4\,\kms\/ (i.e., one tenth of
a resolution element).  From ELF fits to the He~\2 \lam\lam4026, 4200,
4542 absorption, and Si~\4 \lam\lam4098, 4116 emission lines, the mean
heliocentric radial velocity for the first LR02 observations was
$v_{\rm r}$\,$=$\,191.8\,$\pm$1.4 \kms.%\/ (std. err.).

Using the same methods we also cross-correlated the relevant
calibration arcs to investigate the wavelength stability of FLAMES.
The instrument is remarkably stable over all of our observations, with a
largest offset of $\Delta v_{\rm r}$\,$=$\,$-$0.3\,\kms, found between
the arcs for epochs 1 and 6 (i.e., separated by 10 months).

Table~\ref{obsinfo} also lists cross-correlation results for the
regions of the LR03 and 2dF-AAOmega spectra which overlap with the
LR02 data; good agreement is found within the velocity resolution of
the respective data.

\subsubsection{Limits on a binary companion}

To investigate the confidence to which we can rule out a close
massive companion to \#016, we used the method of Sana, Gosset \&
Evans (2009).  In these calculations we adopted a 75\,M$_\odot$
primary, a bi-uniform period ($P$) distribution (with 50\% of the 
simulated systems with 0.3\,$<$\,$\log P$(d)\,$<$\,1 and with 50\% with 
1\,$<$\,$\log P$(d)\,$<$\,3.5), uniform distributions of eccentricities
and mass ratios (between 0 and 0.9, and 0.1 and 1.0, respectively), a 
random orientation in space, and a random periastron passage time.

Our observations allow us to rule out an O- or early
B-type companion (with peak-to-peak variations $>$\,20\,\kms) with a
period in the range 2d\,$<$\,$P$\,$<$\,1yr at the 98.1\% level. These
results are only weakly dependent on the primary mass, with a
variation of $\pm$1\% for a primary in the range of 50--90\,M$_\odot$.

\section{Terminal Wind Velocity}\label{sei}
The Sobolev with exact integration (SEI) method was used to calculate
UV line profiles to estimate $v_\infty$, the terminal wind velocity of
\#016.  We used the {\sc ccp}{\footnotesize 7} SEI
code\footnote{http://ccp7.dur.ac.uk/library.html} which is based on
the method from Lamers, Cerruti-Sola \& Perinotto (1987).  The
observed C~\4 \lam\lam1548-51 doublet was blueshifted by 190\,\kms\/
(Section~\ref{rv}), then compared to the calculated profiles.

The code adopts a standard $\beta$-law to describe the acceleration of
the stellar wind, and includes a turbulence parameter, $v_{\rm t}$
(typically expressed as a fraction of $v_\infty$), to account for
stochastic deviations from the velocity prescription. Line profiles
were calculated over a range of values of $v_\infty$, $\beta$, and $v_{\rm
  t}$, then convolved by the line-spread function noted in 
Section~\ref{obs_uv}.  The best $\chi$-squared fit (excluding the
interstellar absorption features and the emission peak, where the SEI
method is known to underestimate the flux by about 10\%) is obtained
with: $v_\infty$\,$=$\,3450\,$\pm$50\,\kms,
$\beta$\,$=$\,0.75\,$\pm$\,0.1, $v_{\rm
  t}$\,$=$\,0.02\,$\pm$\,0.01\,$v_\infty$. The optical depth law
parameters in the code ($\alpha_{\rm 1}$ and $\alpha_{\rm 2}$) were
set to 1. For saturated lines these can be varied to obtain subtly
different fits to the redward part of the absorption component but
with no significant impact on the derived velocities. 
When $v_{\rm t}$ is relatively small, the value of $v_\infty$
obtained from fitting the line profile effectively corresponds to the blueward
extent (i.e. $v_{\rm black}$) of the saturated P~Cygni profile.

This is one of the highest wind terminal velocities measured 
  directly for an O-type star, behind 
only HDE\,269810 ($v_\infty$\,$=$\,3750\,\kms; Walborn et al. 1995)
and R136a-608\footnote{Star 36 from Massey \& Hunter (1998).}
(O3~If$^\ast$, $v_\infty$\,$=$\,3640\,\kms; Prinja \& Crowther 1998).

\vspace{0.3in}
\section{Quantitative Analysis}

Physical properties for a subset of O2-type stars were presented by
Walborn et al. (2004). We employed similar methods to inform our
discussion of \#016, using {\sc cmfgen} (Hillier \& Miller 1998) 
which solves the radiative transfer equation in the co-moving
frame, under the additional constraint of statistical equilibrium. The
temperature structure follows from the assumption of radiative
equilibrium.  {\sc cmfgen} does not currently solve the momentum
equation, so a density or velocity structure is required. The velocity
for the supersonic part of the wind is parameterized with a classical
$\beta$-type law ($\beta$\,$=$\,0.75, from Section~\ref{sei}), which is
connected to a hydrostatic density structure at depth, such that the
velocity and its gradient match at the interface. The subsonic density
structure is set using a line-blanketed, plane-parallel {\sc tlusty}
model (v.200; Lanz \& Hubeny 2003) with $\log$\,$g$\,$=$\,3.75
(preferred over $\log$\,$g$\,$=$\,4.0). The atomic model is similar to
that adopted by Walborn et al. (2004), including ions from H, He, C,
N, O, Ne, Si, P, S, Ar, Fe and Ni.

We assumed a depth-independent Doppler profile for all lines when
solving for the atmospheric structure in the co-moving frame, and
incoherent electron scattering and Stark broadening for hydrogen and
helium lines was included. A uniform turbulence of 50\,km\,s$^{-1}$
was adopted in the calculation of the emergent spectrum in the
observer's frame. Using the same methods as Walborn et al.  (2010), a
line-broadening parameter ($v\,\sin i$) of 88$\pm$17\,\kms\/ was found
from the optical spectra. Clumping was incorporated using a volume
filling factor, $f$, as described by Hillier et al. (2003), with a
typical value of $f$\,=\,0.1 resulting in a reduction in mass-loss rate by
a factor of $\sqrt{(1/f)} \sim 3$.

\subsection{Spectroscopic results}\label{cmfresults}

Based upon the observed UV to near-IR spectral energy distribution and
the H$\alpha$ and He~\2 \lam\lam1640, 4686 wind features (with
$v_\infty$\,$=$\,3450\,\kms), we estimated the stellar temperature,
luminosity and mass-loss rate of \#016 simultaneously.  

\begin{figure*}
\begin{center}
\includegraphics[scale=0.55,angle=270]{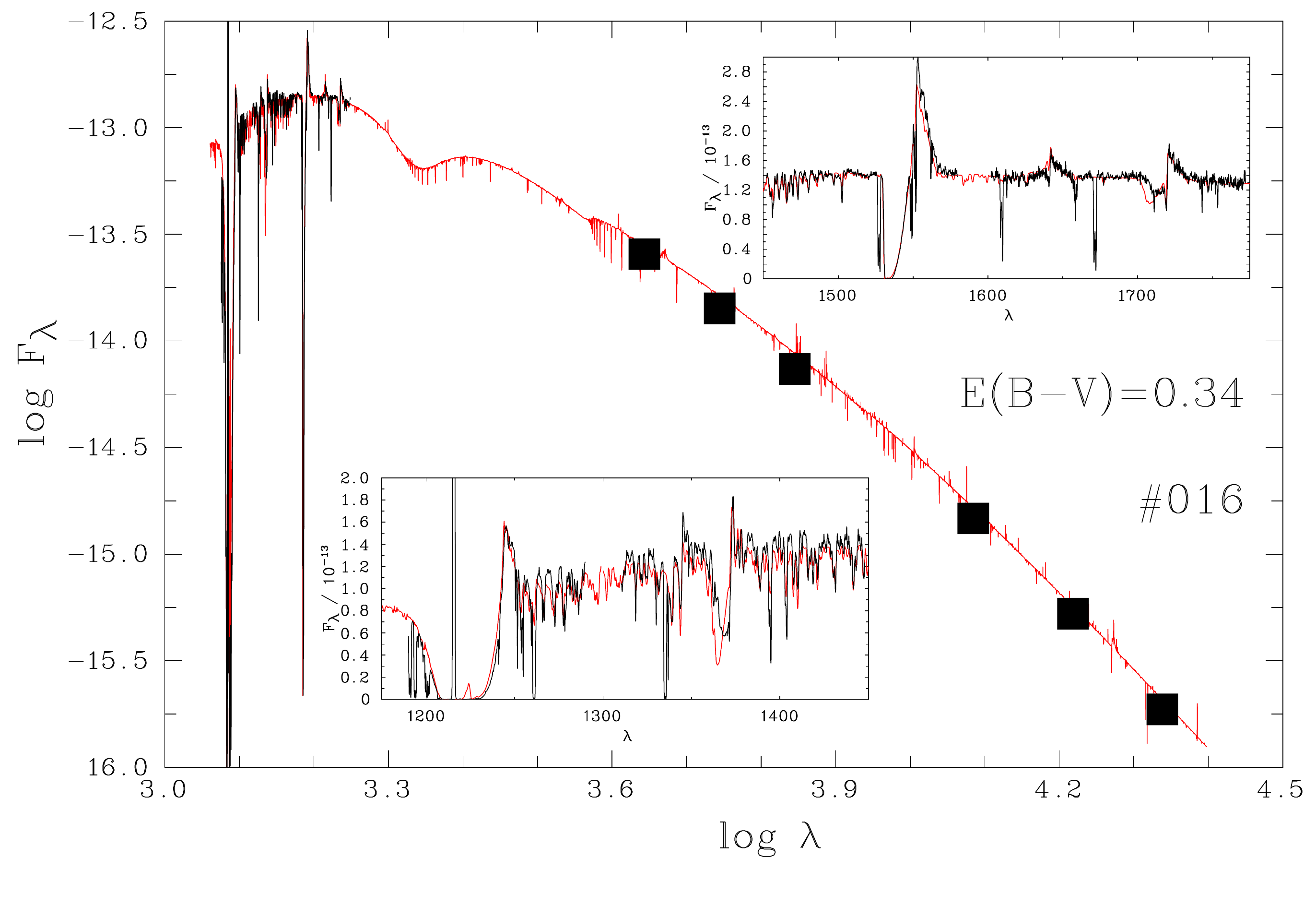}
\caption{{\sc cmfgen} fits (red) to the flux-calibrated UV and
  photometric observations (black lines/squares), used to determine the
  interstellar extinction toward \#016.  Insets are expanded plots of
  the fits to the {\em HST}--COS spectra.}\label{fig3}
\end{center}
\end{figure*}

For an adopted LMC distance of 49\,kpc (distance modulus $=$\,18.45)
the UV spectrophotometry and optical\footnote{$U$\,$=$\,12.62,
  $B$\,$=$\,13.53, $V$\,$=$\,13.49, $R$\,$=$\,13.43 (Massey 2002).}
and near-IR photometry\footnote{$J$\,$=$\,13.38, $H$\,$=$\,13.35,
  $K_{\rm s}$\,$=$\,13.36 (Two Micron All Sky Survey, Skrutskie et al.
  2006)} of \#016 can be reproduced with interstellar extinctions of
$E(B-V)$\,=\,0.07 (foreground) and 0.27 (LMC), using standard Galactic
and LMC (field) extinction laws (Seaton 1979; Howarth 1983) and a
stellar luminosity of
$L_{\ast}$\,$=$\,1.2$\times$\,10$^{6}$\,$L_{\odot}$, as shown in
Figure~\ref{fig3}. The relatively large local extinction is supported
by the strong \lam4428 diffuse interstellar band, which is at a
comparable radial velocity to \#016.

In the absence of measurable He~\1 \lam4471 absorption, a combination
of optical nitrogen lines (N~\4 \lam4058, N~\5 \lam\lam4603-20) and UV
oxygen lines (O~\4 \lam\lam1339-44, O~\5 \lam1371) was used to derive
an estimated effective temperature ($T_{\rm eff}$) of 50\,kK for
\#016. This is a compromise between the available diagnostics --- the
optical N~\5 lines favor a temperature some 5\%--10\% higher, while the
UV O~\4 suggests 10\% lower --- but the strong N~\4 \lam1718 P~Cygni
profile and negligible S~\5 \lam1501 line both firmly support $T_{\rm
  eff}$ = 50\,kK.  Given the adopted wind parameters, a slightly greater
broadening profile was preferred ($v\,\sin i\,\sim$150\,\kms) in the
final model than determined above.

A range of nitrogen abundances were considered, adopting $X_{\rm
  N}$\,=\,0.08\% by mass, which is substantially greater than the
interstellar abundance in the LMC (0.016\%; Russell \& Dopita 1990).
For carbon and oxygen we adopt $X_{\rm C}$\,=\,0.08\% and $X_{\rm
  O}$\,=\,0.2\% by mass, slightly below the results from Dufour (1984)
and Russell \& Dopita (1990) for the interstellar medium (0.095\% and
0.35\%, respectively), although the UV oxygen diagnostics are
relatively insensitive to abundance changes.

The adopted model is shown in Figures~\ref{fig2} and \ref{fig3},
including allowance for an LMC neutral hydrogen column density of
10$^{21.6}$\,cm$^{-2}$.  In general, the fits are satisfactory, with a
few exceptions.  The UV P~Cygni absorption troughs of O~\5 \lam1371
and N~\4 \lam1718 extend too far blueward, although the majority of
the ``iron forest'' (Fe~\5--\6) features are well matched. As with the
temperature, the adopted mass-loss rate ($\dot{M} = 3 \times 10^{-6}
M_{\odot}$ yr$^{-1}$) is a compromise between the available
diagnostics (He~\2 \lam\lam1640, 4686 and H$\alpha$).  Weak metallic
lines in the optical are, in general, reproduced satisfactorily, e.g.,
Si~\4 \lam\lam4089--4116, O~\4 \lam4632, and the blend of C~\4 \lam4658
and O~\4 \lam4654, 4663.  A more detailed treatment of \#016
(including the role of clumping, line-broadening, and the wind
acceleration and turbulence) is beyond the scope of this Letter and
will be presented elsewhere.

A summary of physical and wind parameters of \#016 is provided in
Table~\ref{params}, compared with those for HDE\,269810 from Walborn
et al.  (2004).  We have also compared our results with
LMC-metallicity, non-rotating evolutionary predictions (R.  Hirschi,
2010, private communication) calculated with the mass-loss recipe of
Vink, de~Koter \& Lamers (2001). This approach is supported by our
derived mass-loss rate, which agrees with theoretical predictions to
within $\sim$0.05\,dex for \#016. We obtain the best agreement for a
stellar mass of $\sim$90\,$M_{\odot}$ at an age of $\sim$1\,Myr,
corresponding to $\log$\,$g$\,$\sim$\,4.0. In contrast, the
spectroscopic gravity suggests a considerably lower mass of
$\sim$\,50\,$M_{\odot}$. Using the evolutionary mass to calculate the
escape velocity, $v_{\rm esc}$, our measured terminal velocity is in
excellent agreement with the expected value using the empirical
relation $v_\infty$\,$\sim$\,2.6\,$v_{\rm esc}$ (Lamers, Snow \&
Lindholm 1995), as well as new predictions for the velocity structure
of massive star winds (L. Muijres et al. 2010, in preparation).

\begin{center}
\begin{deluxetable}{cccl}
\tablecaption{Physical and wind properties of 30 Dor 016 
and HDE\,269810\label{params}}
\tablewidth{0pc}
\tablecolumns{3}
\tablehead{\colhead{Star} & \colhead{30 Dor 016} & \colhead{HDE\,269810}}
\startdata
Sp type & O2\,III-If$^{\ast}$   & O2\,III(f$^{\ast}$) \\
$T_{\rm eff}$ (kK) & 50 & 52.5 \\
$\log L/L_{\odot}$ & 6.08    & 6.34 \\
$M_{\odot}$         &  90:      &  130:   \\
$\log$ $g$ (cgs)     & 3.75     & 4.0 \\
$v_{\infty}$ (km\,s$^{-1}$) & 3450 & 3750 \\
$\dot{M}$ ($M_{\odot}$\,yr$^{-1}$) & 10$^{-5.5}$& 10$^{-5.7}$  \\
$M_{\rm V}$ (mag) & --6.0 & --6.6 \\
He/H & $\leq$0.1 & $\leq$0.1  \\
$X_{\rm C}$ (\%) &  0.08 & 0.06  \\
$X_{\rm N}$ (\%) &  0.08  & 0.03  \\
$X_{\rm O}$ (\%) &  0.2  &  0.3 
\enddata
\tablecomments{Mass estimates are obtained from non-rotating
  evolutionary models at LMC metallicity (R. Hirschi, private 
  communication). Metal abundances are reliable to within 50\% 
  (N), or a factor of 2 (C and O).}
\end{deluxetable}
\end{center}

\section{30 Dor 016 as a massive runaway}\label{discussion}

30~Dor~016 is located at the end of an extended rarefied filament of
nebular gas, at a radial distance of $\sim$8\farcm25 (120\,pc in
projection) west--southwest of R136, and 
4\farcm75 (70\,pc) from the less dense cluster NGC\,2060.
The [O~\3] emission superimposed on the stellar spectrum has a peak
velocity of $\sim$275\,\kms, with some asymmetry toward slightly
lower velocities but with no evidence for a second component
matching the velocity of \#016.  This suggests \#016 did not form
locally and is a {\em bona fide} runaway star.  

Further analysis of the stellar and gas dynamics in this region is
warranted to complete our picture of the origins of \#016, e.g., van
Loon \& Zijlstra (2001) found molecular gas to the southeast of R136
at a comparable radial velocity to the star.  At present we lack
proper motions at useful precision to constrain its transverse velocity:
$\mu_\alpha$\,$=$\,1.2\,$\pm$\,5.9, $\mu_\delta$\,$=$\,1.5\,$\pm$\,5.9
mas~yr$^{-1}$ from UCAC2; $\mu_\alpha$\,$=$\,2.8\,$\pm$\,3.8,
$\mu_\delta$\,$=$\,$-$4.3\,$\pm$\,3.8 mas~yr$^{-1}$ from UCAC3 (and
flagged as uncertain; Zacharias et al. 2010).  {\em HST} imaging of
this region would provide much improved constraints (e.g., the
precision achieved by Kallivayalil et al.  2006), and in the longer
term, the {\em Gaia} mission should deliver transverse velocities in 
the LMC to a few \kms.

There are two other O2-type stars (Sk$-$68$^\circ$~137 and BI\,253) to
the north and northwest of 30~Dor, which were proposed as runaways by
Walborn et al. (2002).  BI\,253 has also been observed within the
Tarantula Survey, but from preliminary inspection of its spectra,
its radial velocity appears consistent with this part of the LMC.  The
case of \#016 is also reminiscent of the discovery of N11-026
(O2.5~III(f*)), suggested as a runaway by Evans et al. (2006).

Runaway stars are thought to result from either dynamical interaction
in massive dense clusters, or via a kick from a supernova explosion in
a binary system, with the more massive star exploding first (see,
e.g., Gvaramadze, Gualandris \& Portegies Zwart 2009, and references
therein). It is generally accepted that R136 is sufficiently young
(1-2\,Myr) that its most massive stars have yet to explode as
supernovae.  This implies that, if from R136, \#016 must have been
ejected through dynamical interaction, one of the clearest cases to
date in support of this mechanism. This is vitally important as
dynamical interactions in massive clusters are thought to be a
possible mechanism for producing stellar mergers and very massive
stars (with masses $>$\,140\,$M_\odot$) which might subsequently end
their lives as pair-instability supernovae, of broader relevance to
the early ages of the universe when such massive stars are thought to
be common (e.g., Heger et al. 2003; Gal-Yam et al. 2009).

\section*{Acknowlegements}
We thank Alex de Koter, Paco Najarro, and the referee, Hans Zinnecker,
for their helpful comments.\\

\end{document}